# Developing an Evidence-Based Framework for Grading and Assessment of Predictive Tools for Clinical Decision Support

Mohamed Khalifa[1], Farah Magrabi[1] and Blanca Gallego[1]

[1]*Australian Institute of Health Innovation, Faculty of Medicine and Health Sciences, Macquarie University, Sydney, Australia*

Mohamed Khalifa: mohamed.khalifa@mq.edu.au

Farah Magrabi: farah.magrabi@mq.edu.au

Corresponding Author:

Blanca Gallego: blanca.gallegoluxan@mq.edu.au

**Abstract**

**Background:** Clinical predictive tools quantify contributions of relevant patient characteristics to derive likelihood of diseases or predict clinical outcomes. When selecting a predictive tool, for implementation at clinical practice or for recommendation in clinical guidelines, clinicians are challenged with an overwhelming and ever growing number of tools, most of which have never been implemented or assessed for comparative effectiveness. **Objective:** To develop a comprehensive framework to Grade and Assess Predictive tools (GRASP), and provide clinicians with a standardised, evidence-based system to support their searh for and selection of effective tools. **Methods:** A focused review of literature was conducted to extract criteria along which tools should be evaluated. An initial framework was designed and applied to assess and grade five tools: LACE Index, Centor Score, Well's Criteria, Modified Early Warning Score, and Ottawa knee rule. After peer review, by expert clinicians and healthcare researchers, the framework was revised and the grading of the tools was updated. **Results:** GRASP framework grades predictive tools based on published evidence across three dimensions: 1) Phase of evaluation; 2) Level of evidence; and 3) Direction of evidence. The final grade of a tool is based on the highest phase of evaluation, supported by the highest level of positive evidence, or mixed evidence that supports positive conclusion. Ottawa knee rule had the highest grade since it has demonstrated positive post-implementation impact on healthcare. LACE Index had the lowest grade, having demonstrated only pre-implementation positive predictive performance. **Discussion and Conclusion:** the GRASP framework builds on well-established models and widely accepted concepts to provide standardised assessment and evidence-based grading of predictive tools. Unlike other methods, GRASP is based on the critical appraisal of published evidence reporting the predictive tools' predictive performance before implementation, potential effect and usability during implementation, and their post-implementation impact. Implementing the GRASP framework as an online platform will enable clinicians and clinical guidelines developers to access detailed information, reported evidence and grades of predictive tools. However, keeping the GRASP framework reports up-to-date requires updating tools' assessments and grades when new evidence becomes available. This requires employing semi-automated methods for searching and processing new information.

**Background**

Modern healthcare is building upon information technology to enhance evidence-based practice and cost-effectiveness, using clinical decision support (CDS) systems [1-4]. Based on Shortliffe's definition, there are three levels of CDS: 1) Managing information, 2) Focusing users' attention and 3) Providing patient specific recommendations based on the clinical scenario, which usually follow rules and algorithms, cost benefit analysis or clinical pathways [5, 6]. Clinical predictive tools, here referred to simply as predictive tools, belong to the third level of CDS and include various applications; ranging from the simplest manual clinical prediction rules to the most sophisticated machine learning algorithms [7, 8]. These research-based applications provide diagnostic, prognostic, or therapeutic decision support. They quantify the contributions of relevant patient characteristics to derive the likelihood of diseases, predict their courses and possible outcomes, or support the decision making on their management [9, 10].

*Why do we need grading and assessment of predictive tools?*

When selecting a predictive tool, for implementation at clinical practice or for recommendation in clinical practice guidelines, clinicians involved in the decision making are challenged with an overwhelming and ever growing number of tools. Many of these tools are designed for various clinical contexts and are frequently targeting different patient populations [11-13]. Currently, clinicians rely on their previous experience, subjective evaluation or recent exposure to predictive tools in making selection decisions. Objective methods and evidence-based methods are rarely used in such decisions [14, 15]. Some clinicians, especially those developing clinical guidelines, search the literature for best available evidence, looking for primary studies or systematic reviews on predictive tools. However, there are no available methods to objectively summarise or interpret such evidence [16, 17].

Many clinicians lack the required time and knowledge to evaluate predictive tools, assessing their quality or grading their level of evidence, especially as their number and complexity have increased tremendously in recent years. This is made worse by the complex nature of the evaluation process itself and the variability in the quality of published evidence [18-21]. Although most reported tools have been internally validated, only some have been externally validated and very few have been implemented and studied for their post-implementation impact on healthcare [22, 23]. Various studies and reports indicate that there is an unfortunate practice of developing new tools instead of externally validating or updating existing ones [10, 24-27]. More importantly, while a few pre-implementation studies compare similar predictive tools along some predictive performance measures, comparative studies for post-implementation impact or cost-effectiveness are very rare [28-37]. As a result, there is lack of a reference against which predictive tools can be compared or benchmarked [9, 38, 39].

In addition, decision makers need to consider the usability of a tool, which depends on the specific healthcare and IT settings in which it is embedded, and on users' priorities and perspectives [40]. The usability of tools is consistently improved when the outputs are actionable or directive [41-44]. Moreover, clinicians are keen to know if a tool has been endorsed by certain professional organisations they follow, or recommended by specific clinical guidelines they know [25].

*Current methods for appraising predictive tools*

Several methods have been proposed to evaluate predictive tools [10, 23, 45-59]. However, most of these methods are not based on the critical appraisal of the existing evidence but rather examine a certain aspect of each tool or a specific phase of its development or implementation, reporting evidence about its predictive performance or post-implementation impact. Two exceptions are the TRIPOD statement [60, 61], which provides a set of recommendations for the reporting of studies developing, validating, or updating predictive tools; and the CHARMS checklist [62], which provides guidance on critical appraisal and data extraction for systematic reviews of predictive tools. Both of these methods examine only the pre-implementation predictive performance of the tools, ignoring their usability and post-implementation impact. In addition, none of the currently available methods provide a grading system to allow for benchmarking and comparative effectiveness of tools.

Looking beyond predictive tools, the GRADE framework, grades the quality of published scientific evidence and strength of clinical recommendations, in terms of their post-implementation impact. GRADE has gained a growing consensus, as an objective and consistent method to support the development and evaluation of clinical guidelines, and has increasingly been adopted worldwide [63, 64]. Information based on randomised controlled trials (RCTs) is considered the highest level of evidence. However, the level of evidence could be downgraded due to study limitations, inconsistency of results, indirectness of evidence, imprecision, or reporting bias [64-66]. The strength of a recommendation indicates the extent to which one can be confident that adherence to the recommendation will do more good than harm, it also requires a balance between simplicity and clarity [63, 67].

The aim of this study is to develop a comprehensive evidence-based framework for grading and assessment of predictive tools. This framework is based on the critical appraisal of information provided in the published evidence reporting the evaluation of predictive tools. The framework should provide clinicians with standardised objective information on predictive tools to support their search for and selection of effective tools for their intended tasks. It should support clinicians' informed decision making, whether they are implementing predictive tools at their clinical practices or recommending such tools in clinical practice guidelines to be used by other clinicians.

**Methods**

Guided by the work of Friedman and Wyatt, and their suggested three phases approach, which became an internationally acknowledged standard for evaluating health informatics technologies [19, 20, 40, 68], we aimed to extract the main criteria along which predictive tools can be similarly evaluated before, during and after their implementation.

We started with a focused review of the literature in order to examine and collect the evaluation criteria, and measures proposed for the appraisal of predictive tools along these three phases. The concepts used in the search included "clinical prediction", "tools", "rules", "models", "algorithms", "evaluation", and "methods". The search was conducted using four databases; MEDLINE, EMBASE, CINAHL and Google Scholar, with no specific timeframe. This literature review was then extended to include studies describing methods evaluating CDS systems and more generally, health information systems and technology. Following the general concepts of the PRISMA guilines [69], the duplicates of the retrieved studies, from the four databases, were first removed. Studies were then screened, based on their titles and abstracts, for relevance, then the full text articles were assessed for eligibility and only the eligible studies were included in the review. We included three types of studies evaluating predictive tools and other CDS systems; 1) studies describing the methods or processes of the evaluation, 2) studies describing the phases of the evaluation, and 3) studies describing the criteria and measures used in the evaluation. The first author manually extracted the methods of evaluation described in each type of study, and this was then revised and confirmed by the second and last authors. Figure 3, in the Appendix, shows the process of study selection for inclusion in the focused review of the literature.

Using the extracted information, we designed an initial version of the framework and applied it to asses and grade five predictive tools. We reviewed the complete list of 426 tools published by the MDCalc medical reference website for decision support tools and applications and calculators (https://www.mdcalc.com) [70]. We excluded tools which are not clinical - their output is not related to providing individual patient care, such as scores of ED crowding and calculators of waiting times. We also excluded tools which are not predictive - their output is not the result of statistically generating new information but rather the result of a deterministic equation, such as calculators deriving a number from laboratory results. The five example tools were then randomly selected, using a random number generator [71], from a shorter list of 107 eligible predictive tools, after being alphabetically sorted and numbered.

A comprehensive and systematic search for the published evidence, on each of the five predictive tools, was conducted, using MEDLINE, EMBASE, CINAHL and Google Scholar, and refined in four steps. 1) The primary studies, describing the development of the tools, were first identified and retrieved. 2) All secondary studies that cited the primary studies or that referred to the tools' names or to any of their authors, anywhere in the text, were retrieved. 3) All tertiary studies that cited the secondary studies or that were used as references by the secondary studies were retrieved. 4) Secondary and tertiary studies were examined to exclude non-relevant studies or those not reporting the validation, implementation or evaluation of the tools. After the four steps, eligible evidence was examined and grades were assigned to the predictive tools. Basic information about the tool, such as year of publication, intended use, target population, target outcome, and source and type of input data were extracted, from the primary studies, to inform the first "Tool Information" section of the framework. Eligible studies were then examined in detail for the reported evaluations of the predictive tools. Figure 4, in the Appendix, shows the process of searching the literature for the published evidence on the predictive tools.

The framework and its application to the selected five predictive tools were then peer reviewed by six expert healthcare professionals. Three of these professionals are clinicians, who work in hospitals and have over ten years of experience using CDS systems, while the other three are healthcare researchers, who work in research organisations and have over twenty years of experience in developing, implementing or evaluating CDS systems. The reviewers were provided with the framework's concept design and its detailed report template. They were also provided with the summarised and detailed grading of the five predictive tools, as well as the justification and published evidence underpinning the grade assignment. After a brief orientation session, reviewers were asked to feedback on how much they agreed with each of the framework's dimensions and corresponding evaluation criteria, the 'Tool information' section as well as the grading of the five exemplar predictive tools. The framework was then refined and the grading of the five predictive tools was updated based on the reviewers' feedback. Figure 1 shows the flowchart of the GRASP framework overall development process.

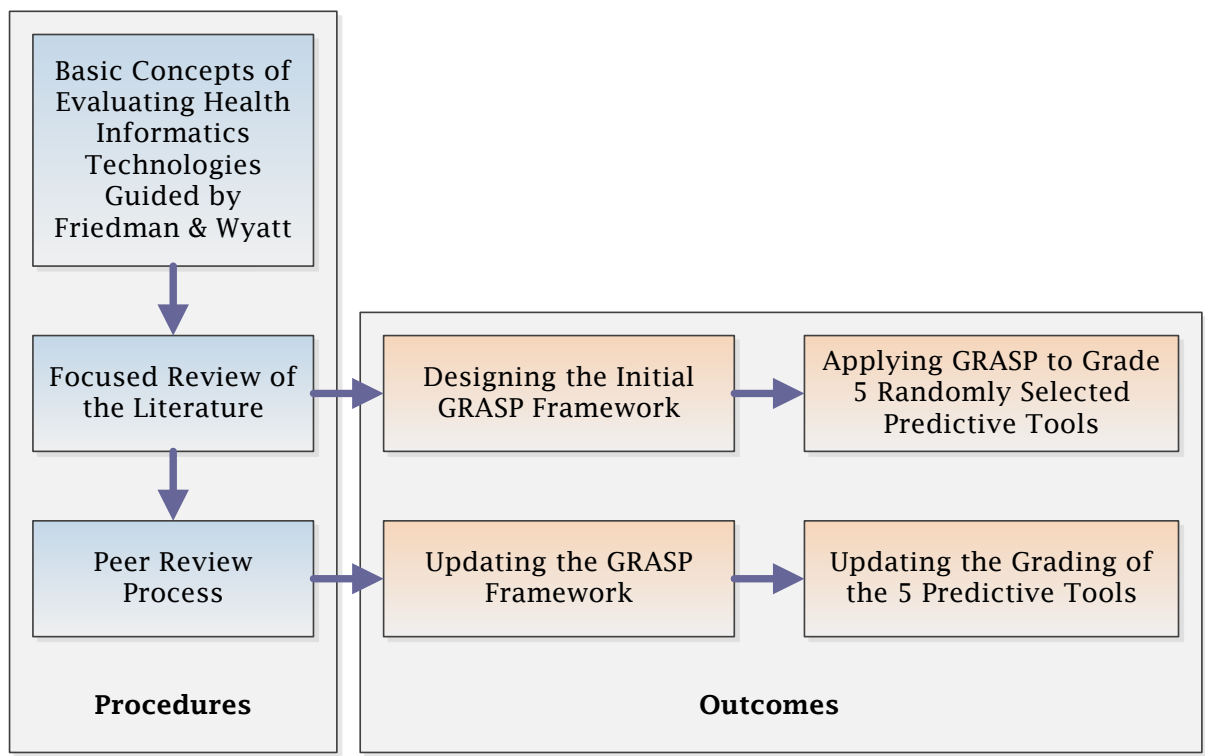

Figure 1: The GRASP Framework Development Process

## Results

*The Focused Review of Literature*

The search in the four databases, after removing the duplicates, identified a total of 831 studies. After screening the titles and abtracts, 647 studies were found not relevant to the topic. The full text of the remaining 184 studies were then examined to exclude non-eligible studies, which were 134 studies, based on the inclusion criteria. Only fifty studies were identified as eligible. Twenty three of the fifty studies described methods for the evaluation of predictive tools [10, 12, 23, 39, 43, 45-60, 62, 72], ten studies described the evaluation of CDS systems [2, 73-81], and eleven studies described the evaluation of hospital information systems and technology [1, 82-91]. One study described the NASSS framework, a guideline to help predict and evaluate the success of healthcare technologies [92]; and five studies described the GRADE framework for evaluating clinical guidelines and protocols [63-67]. The following three subsections describe the methods used to evaluate predictive tools as described in the focussed literature review. A summary of the evaluation criteria and examples of corresponding measures for each phase can be found in Table 3 in the Appendix.

*Before Implementation – Predictive performance*

During the development phase, the internal validation of the predictive performance of a tool is the first step to make sure that the tool is doing what it is intended to do [53, 54]. Predictive performance is defined as the ability of the tool to utilise clinical and other relevant patient variables to produce an outcome that can be used to supports diagnostic, prognostic or therapeutic decisions made by clinicians and other healthcare professionals [9, 10]. The predictive performance of a tool is evaluated using measures of discrimination and calibration [52]. Discrimination refers to the ability of the tool to distinguish between patients with and without the outcome under consideration. This can be quantified with measures such as sensitivity, specificity, and the area under the receiver operating characteristic curve – AUC (or concordance statistic, c). The D-statistic is a measure of discrimination for time-to-event outcomes, which is commonly used in validating the predictive performance of prognostic models using survival data [93]. The log-rank test, or sometimes referred to as the Mantel-Cox test, is used to establish if the survival distributions of two samples of patients are statistically different. They are commonly used to validate the discrimination power of clinical prognostic models [94]. On the other hand, calibration refers to the accuracy of prediction, and indicates the extent to which expected and observed outcomes agree [47, 55]. Calibration is measured by plotting the observed outcome rates against their corresponding predicted probabilities. This is usually presented graphically with a calibration plot that shows a calibration line, which can be described with a slope and an intercept [95]. It is sometimes summarised using the Hosmer-Lemeshow test or the Brier score [96]. To avoid over-fitting, tools' predictive performance must always be assessed out-of-sample, either via cross-validation or bootstrapping [55]. Of more interest than the internal validity is the external validity (reliability or generalisability), where the predictive performance of a tool is estimated in independent validation samples of patients from different populations [51].

*During Implementation – Potential Effect & Usability*

Before wide implementation, it is important to learn about the estimated potential effect of a predictive tool, when used in the clinical practice, on three main categories of measures: 1) Clinical effectiveness, such as improving patient outcomes, estimated through clinical effectiveness studies, 2) Healthcare efficiency, including saving costs and resources, estimated through feasibility and cost-effectiveness studies, and 3) Patient safety, including minimising complications, side effects, and medical errors. These categories are defined by the Institute of Medicine as objectives for improving healthcare performance and outcomes, and are differently prioritised by clinicians, healthcare professionals and health administrators [97, 98]. The potential effect, of a predictive tool, is defined as the expected, estimated or calculated impact of using the tool on different healthcare aspects, processes or outcomes, assuming the tool has been successfully implemented and is used in the clinical practice, as designed by its developers [40, 99]. A few predictive tools have been studied for their potential to enhance clinical effectiveness and improve patient outcomes. For example, the spinal manipulation clinical prediction rule was tested, before implementation, on a small sample of patients to identify those with low back pain most likely to benefit from spinal manipulation [100]. Other tools have been studied for their potential to improve healthcare efficiency and save costs. For example, using a decision analysis model, and assuming all eligible children with minor blunt head trauma were managed using the CHALICE rule (Children's Head Injury Algorithm for the Prediction of Important Clinical Events), it was estimated that CHALICE would reduce unnecessary expensive head computed tomography (CT) scans, by 20%, without risking patients' health [101-103]. Similarly, the use of the PECARN (Paediatric Emergency Care Applied Research Network) head injury rule was estimated to potentially improve patient safety through minimising the exposure of children to ionising radiation resulting in fewer radiation-induced cancers and lower net quality adjusted life years loss [104, 105].

In addition, it is important to learn about the usability of predictive tools. Usability is defined as the extent to which a system can be used by the specified users to achieve specified and quantifiable objectives in a specified context of use [106, 107]. There are several methods to make a system more usable and many definitions have been developed, based on the perspective of what usability is and how it can be evaluated, such as the mental effort needed and the user attitude or the user interaction, represented in the easiness of use and acceptability of systems [108, 109]. Usability can be evaluated through measuring the effectiveness of task management with accuracy and completeness, measuring efficiency of utilising resources in completing tasks and measuring users' satisfaction, comfort with, and positive attitudes towards, the use of the tools [110, 111]. More advanced techniques, such as think aloud protocols and near live simulations, are recently used to evaluate usability [112]. Think aloud protocols are a major method in usability testing, since they produce a larger set of information and a richer content. They are conducted either retrospectively or concurrently, where each method has its own way of detecting usability problems [113]. The near live simulations provide users, during testing, with an opportunity to go through different clinical scenarios while the system captures interaction challenges and usability problems [114, 115]. Some researchers add learnability, memorability and freedom of errors to the measures of usability. Learnability is an important aspect of usability and a major concern in the design of complex sysems. It is the capability of a system to enable the users to learn how to use it. Momerability, on the other hand, is the capability of a system to enable the users to remember how to use it, when they return back. Learnability and memorability are measured through subjective survey methods, asking users about their experience after using systems, and can also be measured by monitoring users' competence and learning curves over successive sessions of system usage [116, 117].

*After Implementation – Post-Implementation Impact*

Some predictive tools have been implemented and used in the clinical practice for years, such as the PECARN head injury rule or the Ottawa knee and ankle rules [118-120]. In such cases, clinicians might be interested to learn about their post-implementation impact. The post-implementation impact of predictive tools is defined as the achieved change or influence, of a predictive tool, on different healthcare aspects, processes or outcomes, after the tool has been successfully implemented and used in the clinical practice, as designed by its developers [2, 41]. Similar to the measures of potential effect, post-implementation impact is reported along three main categories of measures: 1) Clinical effectiveness, such as improving patient outcomes, 2) Healthcare efficiency, such as saving costs and resources, and 3) Patient safety, such as minimising complications, side effects, and medical errors. These three categories of post-implementation impact measures are differently prioritised by clinicians, healthcare professionals and health administrators. In this phase of evaluation, we follow the main concepts of the GRADE framework, where the level of evidence for a given outcome is firstly determined by the study design [63, 64, 67]. High quality experimental studies, such as randomised and nonrandomised controlled trials, and the systematic reviews of their findings, come on top of the evidence levels followed by observational well-designed cohort or case-control studies and lastly subjective studies, opinions of respected authorities, and reported of expert committees or panels [64-66]. For simplicity, we did not include GRADE's detailed criteria for higher and lower quality of studies. However, effect sizes and potential biases are reported as part of the framework, so that consistency of findings, trade-offs between benefits and harms, and other considerations can also be assessed.

*Developing the GRASP Framework*

Our suggested GRASP framework (abbreviated from Grading and Assessment of Predictive Tools) is illustrated in Table 1. Based on published evidence, the GRASP framework uses three dimensions to grade predictive tools: **Phase of Evaluation**, **Level of Evidence** and **Direction of Evidence**.

*Phase of Evaluation:* Assigns A, B and/or C based on the highest phase of evaluation. If a tool's predictive performance, as reported in the literature, has been tested for validity, it is assigned phase C. If a tool's usability and/or potential effect have been tested, it is assigned phase B. Finally, if a tool has been implemented in clinical practice, and there is published evidence evaluating its post-implementation impact, it is assigned phase A.

*Level of Evidence:* A numerical score, within each phase, is given based on the level of evidence associated with each tool. A tool is graded C1 if it has been tested for external validity multiple times; C2 if it has been tested for external validity only once; and C3 if it has been only tested for internal validity. Corresponding measures of discrimination and calibration, as reported in the literature, are listed; together with additional relevant information such as if the study evaluated the tool in a subpopulation of the tool's intended target population. Similarly, B1 is assigned to a predictive tool that has been evaluated during implementation for its usability; while if it has been studied for its potential effect on clinical effectiveness, patient safety or healthcare efficiency, it is assigned B2. Finally, if a predictive tool had been implemented then evaluated after implementation for its post-implementation impact, it is assigned score A1 if there is at least one experimental study evaluating its post-implementation impact, A2 if there are observational studies evaluating its post-implementation impact and A3 if the post-implementation impact has been evaluated only through subjective or descriptive studies. Effect sizes for each outcome of interest together with study type, clinical settings and patient subpopulations are reported.

*Direction of Evidence:* Due to the large heterogeneity in study design, outcome measures and patient subpopulations contained in the studies, synthesising measures of predictive performance, usability, potential effect or post-implementation impact into one quantitative value is not possible. Furthermore, acceptable values of predictive performance or post-implementation impact measures depend on the clinical context and the task at hand. For example, tools like Ottawa knee rule [120] and Wells' criteria [121, 122] are considered effective only when their sensitivity is very close to 100%, since their task is to identify patients with fractures or pulmonary embolism before sending them home. On the other hand, tools like LACE Index [123] and Centor score [124] are accepted to show sensitivities of around 70%, since their tasks, to predict 30 days readmission risk or identify that pharyngitis is bacterial, aim to screen patients who may benefit from further interventions. Therefore, for each phase and level of evidence, we assign a direction of evidence, based on the conclusions reported in the studies, and provide the user with the option to look at the summary of the findings for further information. The evidence is considered positive if all studies about a predictive tool reported positive conclusions and negative if all studies reported negative or equivocal conclusions. The evidence is considered mixed if some studies reported positive and some reported either negative or equivocal conclusions. Evaluating the evidence direction based on the conclusions of studies is shown in detail in Table 4 in the Appendix.

To decide an overall direction of evidence, we developed a protocol to sort the mixed evidence into 1) Mixed evidence that supports an overall positive conclusion or 2) Mixed evidence that supports an overall negative conclusion. This protocol is based on two main criteria; 1) Degree of matching between the evaluation study conditions and the original tool specifications, and 2) Quality of the evaluation study. Studies evaluating predictive tools in closely matching conditions to the tool specifications and providing high quality evidence are considered first; taking into account their conclusions in deciding the overall direction of evidence. The mixed evidence protocol is detailed and illustrated in Figure 5 in the Appendix.

The final grade of a predictive tool is based on the highest phase of evaluation, supported by the highest level of positive evidence, or mixed evidence that supports an overall positive conclusion. Figure 2 shows the GRASP framework concept; a visual presentation of the framework three dimensions, phase of evaluation, level of evidence, and direction of evidence, explaining how each tool is assigned the final grade. Table 1 shows the GRASP framework detailed report.

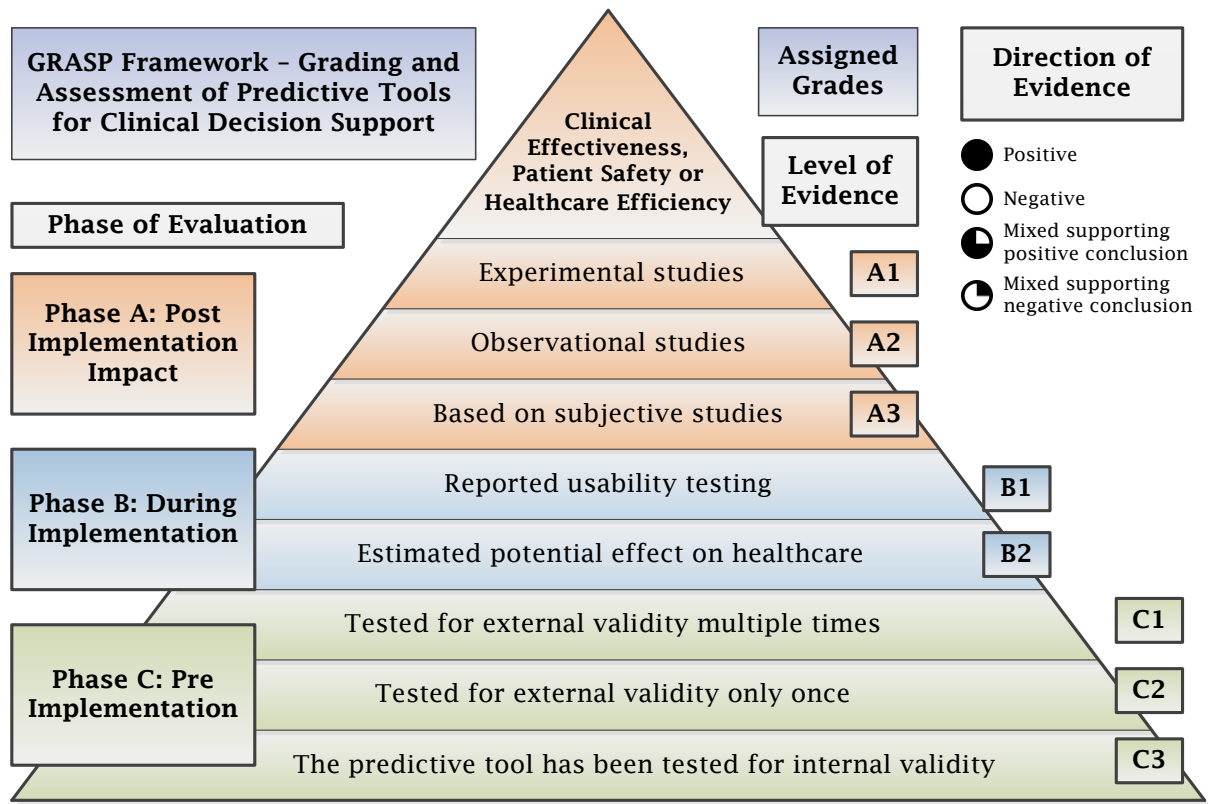

Figure 2: The GRASP Framework Concept

Table 1: The GRASP Framework Detailed Report

| Name | Name of predictive tool (report tool's creators and year in the absence of a given name) | | |
|---|---|---|---|
| Authors/Year | Name of developer, country and year of publication | | |
| Intended use | Specific aim/intended use of the predictive tool | | |
| Intended user | Type of practitioner intended to use the tool | | |
| Category | Diagnostic/Therapeutic/Prognostic/Preventive | | |
| Clinical area | Clinical specialty | | |
| Target Population | Target patient population and health care settings in which the tool is applied | | |
| Target Outcome | Event to be predicted (including prediction lead time if needed) | | |
| Action | Recommended action based on tool's output | | |
| Input source | • Clinical (including Diagnostic, Genetic, Vital signs, Pathology)<br>• Non-Clinical (including Healthcare Utilisation) | | |
| Input type | • Objective (Measured input; from electronic systems or clinical examination)<br>• Subjective (Patient reported; history, checklist …etc.) | | |
| Local context | Is the tool developed using location-specific data? (e.g. life expectancy tables) | | |
| Methodology | Type of algorithm (e.g. parametric/non-parametric) | | |
| Endorsement | Organisations endorsing the tool and/or guidelines recommending its utilisation | | |
| Automation Flag | Automation status (manual/automated) | | |
| Tool Citations | Total citations of the tool | Number of studies reporting the tool | |
| Phase of Evaluation | Level of Evidence | Grade | Evaluation Studies |
| Phase C: | Insufficient internal validation | C0 | Tested for internally validity but was either insufficiently internally validated or validation was insufficiently reported. |

| Before implementation **Is it possible?** | Internal validation | C3 | Tested for internally validity (reported calibration & discrimination; sensitivity, specificity, positive and negative predictive values & other predictive performance measures). | | | | |
|---|---|---|---|---|---|---|---|
| | External validation | C2 | Tested for external validity, using one external dataset. | | | | |
| | External validation multiple times | C1 | Tested multiple times for external validity, using more than one external dataset. | | | | |
| **Phase B:** During implementation **Is it practicable?** | Potential effect | B2 | Reported estimated potential effect on clinical effectiveness, patient safety or healthcare efficiency. | | | | |
| | Usability | B1 | Reported usability testing (effectiveness, efficiency, satisfaction, learnability, memorability, and minimizing errors). | | | | |
| **Phase A:** After implementation: **Is it desirable?** | Evaluation of Post-Implementation Impact on Clinical Effectiveness, Patient Safety or Healthcare Efficiency | A3 | Based on subjective studies; e.g. the opinion of a respected authority, clinical experience, a descriptive study, or a report of an expert committee or panel. | | | | |
| | | A2 | Based on observational studies; e.g. a well-designed cohort or case-control study. | | | | |
| | | A1 | Based on experimental studies; e.g. a well-designed, widely applied randomised/nonrandomised controlled trial. | | | | |
| **Final Grade** | Grade ABC,123 | A1 | A2 | A3 | B1 | B2 | C1 | C2 | C3 |
| **Direction of Evidence** | ● Positive Evidence | ◐ Mixed Evidence Supporting Positive Conclusion | | | | | | |
| | ○ Negative Evidence | ◑ Mixed Evidence Supporting Negative Conclusion | | | | | | |
| **Justification** | Explains how the final grade is assigned based on evidence; which conclusions were taken into consideration, as positive evidence, and which were considered negative. | | | | | | | |
| **References** | Details of studies that support the justification: phase of evaluation, level of evidence, direction of evidence, study type, study settings, methodology, results, findings and conclusions (highlighted according to the colour code). | These two sections are included in the full GRASP report on each tool. | | | | | | |
| **Label/Colour Code** | • Positive Findings • Negative Findings | • Important Findings • Less Relevant Findings | | | | | | |

*Applying the GRASP Framework to Grade Five Predictive Tools*

In order to show how GRASP works, we applied it to grade five randomly selected predictive tools; LACE Index for Readmission [123], Centor Score for Streptococcal Pharyngitis [124], Wells' Criteria for Pulmonary Embolism [121, 122, 125], The Modified Early Warning Score (MEWS) for Clinical Deterioration [126] and Ottawa Knee Rule [120]. In addition to these seven primary studies, describing the development of the five predictive tools, our systematic search for the published evidence revealed a total of fifty six studies; validating, implementing, and evaluating the five predictive tools. The LACE Index was evaluated and reported in six studies, the Centor Score in fourteen studies, the Wells' Criteria in ten studies, the MEWS in twelve studies, and the Ottawa Knee Rule in fourteen studies. To apply the GRASP framework and assign a grade to each predictive tool, the following steps were conducted; 1) The primary study or studies were first examined for the basic information about the tool and the reported details of development and validation. 2) Other studies were examined for their phases of evaluation, levels of evidence and direction of evidence. 3) Mixed evidence was sorted into positive or negative. 4) The final grade was assigned and supported by the detailed justification. A summary of grading the five tools is shown in Table 2 and a detailed GRASP report on each tool is provided in the Appendix; Tables 5 to 9.

**LACE Index** is a prognostic tool designed to predict 30 days readmission or death of patients after discharge from hospitals. It uses multivariable logistic regression analysis of four administrative data elements; length of stay, admission acuity, comorbidity (Charlson Comorbidity Index) and emergency department (ED) visits in the last six months, to produce a risk score [123]. The tool has been tested for external validity twice; using a sample of 26,045 patients from six hospitals in Toronto and a sample of 59,652 patients from all hospitals in Alberta, Canada. In both studies, the LACE Index showed positive external validity and superior predictive performance to the previous similar tools endorsed by the Centres for Medicare and Medicaid Services in the United States [127, 128].

Two studies examined the predictive performance of LACE Index on small sub-population samples; 507 geriatric patients in the United Kingdom and 253 congestive heart failure patients in the United States, and found that the index performed poorly [129, 130]. Two more studies reported that the LACE Index performed well but not better that their own developed tools [131, 132]. Using the mixed evidence protocol, the mixed evidence here supports external validity, since the two negative conclusion studies have been conducted on very small samples of patients and on different subpopulations than the one the LACE Index was developed for. There was no published evidence on the usability, potential effect or post-implementation impact of the LACE Index. Accordingly, the LACE Index has been assigned Grade C1.

**Centor Score** is a diagnostic tool that uses a rule-based algorithm on clinical data to estimate the probability that pharyngitis is streptococcal in adults who present to the ED complaining of sore throat [124]. The score has been tested for external validity multiple times and all the studies reported positive conclusions [133-140]. This qualifies Centor score for Grade C1. One study conducted a multicentre cluster RCT usability testing of the integration of Centor score into electronic health records. The study used "Think Aloud" testing with ten primary care providers, post interaction surveys in addition to screen captures and audio recordings to evaluate usability. Within the same study, another "Near Live" testing, with eight primary care providers, was conducted. Conclusions reported positive usability of the tool and positive feedback of users on the easiness of use and usefulness [141]. This qualifies Centor score for Grade B1.

Evidence of the post-implementation impact of Centor score is mixed. One RCT conducted in Canada reported a clinically important 22% reduction in overall antibiotic prescribing [142]. Four other studies, three of which were RCTs, reported that implementing Centor score did not reduce antibiotic prescribing in clinical practice [143-146]. Using the mixed evidence protocol, we found that the mixed evidence does not support positive post-implementation impact of Centor score. Therefore, Centor score has been assigned Grade of B1.

**Wells' Criteria** is a diagnostic tool used in the ED to estimate pre-test probability of pulmonary embolism [121, 122]. Using a rule-based algorithm on clinical data, the tool calculates a score that excludes pulmonary embolism without diagnostic imaging [125]. The tool was tested for external validity multiple times [147-151] and its predictive performance has been also compared to other predictive tools [152-154]. In all studies, Wells' criteria was reported externally valid, which qualifies it for Grade C1. One study conducted usability testing for the integration of the tool into the electronic health record system of a tertiary care centre's ED. The study identified a strong desire for the tool and received positive feedback on the usefulness of the tool itself. Subjects responded that they felt the tool was helpful, organized, and did not compromise clinical judgment [155]. This qualifies Wells' criteria for Grade B1. The post-implementation impact of Well's Criteria on efficiency of computed tomography pulmonary angiography (CTPA) utilisation has been evaluated through an observational before-and-after intervention study. It was found that the Well's Criteria significantly increased the efficiency of CTPA utilisation and decreased the proportion of inappropriate scans [156]. Therefore, Well's Criteria has been assigned Grade A2.

**The Modified Early Warning Score (MEWS)** is a prognostic tool for early detection of inpatients' clinical deterioration and potential need for higher levels of care. The tool uses a rule-based algorithm on clinical data to calculate a risk score [126]. The MEWS has been tested for external validity multiple times in different clinical areas, settings and populations [157-163]. All studies reported that the tool is externally valid. However, one study reported MEWS poorly predicted the in-hospital mortality risk of patients with sepsis [164]. Using the mixed evidence protocol, the mixed evidence supports external validity, qualifying MEWS for Grade C1. No literature has been found regarding its usability or potential effect.

The MEWS has been implemented in different healthcare settings. One observational before-and-after intervention study failed to prove positive post-implementation impact of the MEWS on patient safety in acute medical admissions [165]. However, three more recent observational before-and-after intervention studies reported positive post-implementation impact of the MEWS on patient safety. One study reported significant increase in frequency of patient observation and decrease in serious adverse events after intensive care unit (ICU) discharge [166]. The second reported significant increase in frequency of vital signs recording, 24h post-ICU discharge and 24h preceding unplanned ICU admission [167]. The third, an eight years study, reported that the post-implementation four years showed significant reductions in the incidence of cardiac arrests, the proportion of patients admitted to ICU and their in-hospital mortality [168]. Using the mixed evidence protocol, the mixed evidence supports positive post-implementation impact. The MEWS has been assigned Grade A2.

**Ottawa Knee Rule** is a diagnostic tool used to exclude the need for an X-ray for possible bone fracture in patients presenting to the ED, using a simple five items manual check list [120]. It is one of the oldest, most accepted and successfully used rules in CDS. The tool has been tested for external validity multiple times. One systematic review identified 11 studies, 6 of them involved 4,249 adult patients and were appropriate for pooled analysis, showing high sensitivity and specificity predictive performance [169]. Furthermore, two studies discussed the post-implementation impact of Ottawa knee rule on healthcare efficiency. One nonrandomised controlled trial with before-after and concurrent controls included a total of 3,907 patients seen during two 12-month periods before and after the intervention. The study reported that the rule decreased the use of knee radiography without patient dissatisfaction or missed fractures and was associated with reduced waiting times and costs per patient [170]. Another nonrandomised controlled trial reported that the proportion of ED patients referred for knee radiography was reduced. The study also reported that the practice based on the rule was associated with significant cost savings [171]. Accordingly, the Ottawa knee rule has been assigned Grade A1.

In the Appendix, a summary of the predictive performance of the five tools is shown in Table 10. The c-statistics of LACE Index, Centor Score, Wells' Criteria and MEWS are reported in Figure 6. The usability of Centor Score and Wells Criteria are reported in Table 11 and post-implementation impact of Wells Criteria, MEWS and Ottawa knee rule is reported in Table 12.

Table 2: Summary of Grading the Five Predictive Tools

| Tool Name | Tool Information | | | | Tool Grade | Impact After Implementation | | | During Implementation | | Predictive performance Before Implementation | | |
|---|---|---|---|---|---|---|---|---|---|---|---|---|---|
| | Country | Year | Citations | Studies | | Experimental Studies | Observational Studies | Subjective Studies | Usability | Potential Effect | External Validation Multiple Times | External Validation Only Once | Internal Validation |
| | | | | | | A1 | A2 | A3 | B1 | B2 | C1 | C2 | C3 |
| LACE Index [123] | Canada | 2010 | 455 | 7 | C1 | | | | | | ◐ | | ● |
| Centor Score [124] | USA | 1981 | 715 | 15 | B1 | ◐ | | | ● | | ● | | ● |
| Wells' Criteria [121, 122, 125] | Canada | 1998 | 1,260 | 13 | A2 | | ● | | ● | | ● | | ● |
| Modified Early Warning Score [126] | UK | 2001 | 1,176 | 13 | A2 | | ◐ | | | | ◐ | | ● |
| Ottawa Knee Rule [120] | Canada | 1995 | 227 | 15 | A1 | ● | | | | | ● | | ● |
| Evidence Direction | ● Positive Evidence | | | | | ◐ Mixed Evidence Supporting Positive Conclusion | | | | | | | |
| | ○ Negative Evidence | | | | | ◐ Mixed Evidence Supporting Negative Conclusion | | | | | | | |

*Peer Review of the GRASP Framework*

On peer-review, experts found the GRASP framework logical, helpful and easy to use. The reviewers strongly agreed to all criteria used for evaluation. The reviewers suggested adding more specific information about each tool, such as the author's name, the intended user of the tool and the recommended action based on the tool's findings. The reviewers showed a clear demand for knowledge regarding the applicability of tools to their local context. Two aspects were identified and discussed with the reviewers. Firstly, the operational aspect of how easy it would be to implement a particular tool and if the data required to use the tool is readily available in their clinical setting. Secondly, the validation aspect of adopting a tool developed using local predictors, such as life expectancy (which is location specific) or information based on billing codes (which is hospital specific). Following this discussion, elements related to data sources and context were added to the information section of the framework. One reviewer suggested assigning grade C0 to the reported predictive tools that did not meet C3 criteria, i.e. those tools which were tested for internally validity but were either insufficiently internally validated or the internal validation was insufficiently reported in the study, in order to differentiate them from those tools for which neither predictive performance nor post-implementation impact have been reported in the literature.

**Discussion and Conclusion**

*Brief Summary*

It is challenging for clinicians to critically evaluate the growing number of predictive tools proposed to them by colleagues, administrators and commercial entities. Although most of these tools have been assessed for predictive performance, only a few have been implemented or evaluated for comparative predictive performance or post-implementation impact. In this manuscript, we present GRASP, a framework that provides clinicians with an objective, evidence-based, standardised method to use in their search for, and selection of tools. GRASP builds on widely accepted concepts, such as Friedman and Wyatt's evaluation approach [19, 20, 40, 68] and the GRADE system [63-67].

The GRASP framework is designed for two levels of users: 1) Expert users, who will use the framework to assign grades to predictive tools and report their details, through the critical appraisal of published evidence about these tools. Expert users include healthcare researchers who specialise in evidence-based methods and have experience in developing, implementing or evaluating predictive tools. 2) End users, who will use the GRASP framework detailed report of tools and their final grades, produced by expert users, to compare existing predictive tools and chose the most suitable for their tasks and settings. End users include clinicians and other healthcare professionals involved in the decision making and the selection of predictive tools for implementation at their clinical practice or for recommendation in clinical practice guidelines to be used by other clinicians and healthcare professionals.

*Comparison with Previous Literature*

Previous approaches to the appraisal of predictive tools from the published literature, namely the TRIPOD statement [60, 61] and the CHARMS checklist [62], examine only their predictive performance, ignoring their usability and post-implementation impact. On the other hand, the GRADE framework appraises the published literature in order to evaluate clinical recommendations based on their post-implementation impact [63-67]. More broadly, methods for the evaluation of health information systems and technology focus on the integration of systems into tasks, workflows and organisations [83, 89]. The GRASP framework takes into account all phases of the development and translation of a predictive algorithm: predictive performance before implementation, using similar concepts as those utilised in TRIPOD and CHARMS; usability and potential effect during implementation, and post-implementation impact on patient outcomes and processes of care after implementation, using similar concepts as those utilised in the GRADE system. The GRASP grade is not the result of combining and synthesising selected measures of predictive performance (e.g. AUC), potential effect (e.g. potential saved money), usability (e.g. user satisfaction) or post-implementation impact (e.g. increased efficacy) from the existing literature, like in a meta-analysis; but rather the result of combining and synthesising the reported qualitative conclusions.

Walker and Habboushe, at the MDCalc website; classified and reported the most commonly used medical calculators and other clinical decision support applications. However, the website does not provide users with a structured grading system or an evidence-based method for the assessment of the presented tools. Therefore, we believe that our proposed framework can be adopted and used by MDCalc, and similar clinical decision support resources, to grade their tools.

*Quality of Evidence and Conflicting Conclusions*

One of the main challenges is dealing with the large variability in the quality and type of studies in the published literature. This heterogeneity makes it impractical to synthesise measures of predictive performance, usability, potential effect or post-implementation impact into single numbers. Furthermore, as discussed earlier, a particular value of a predictive performance metric that is considered good for some tasks and clinical settings may be considered insufficient for others. In order to avoid complex decisions regarding the quality and strength of reported measures, we chose to assign a direction of evidence, based on positive or negative conclusions as reported in the studies under consideration. We then provide the end user with the option to look at a summary of reported measures for further details.

It is not uncommon to encounter conflicting conclusions when a tool has been validated in different patient subpopulations. For example, LACE Index for readmission showed positive external validity when tested in adult medical inpatients [127, 128], but showed poor predictive performance when tested in a geriatric subpopulation [129]. Similarly, the MEWS for clinical deterioration demonstrated positive external validity when tested in emergency patients [157, 159, 163], medical inpatients [158, 162], surgical inpatients [160], and trauma patients [161], but not when tested in a subpopulation of patients with acute sepsis [164]. Part of these disagreements could be explained by changes in the distributions of important predictors, which affect the discriminatory power of the algorithms. For example, sepsis patients have similarly disturbed physiological measures such as those used to generate MEWS. In addition, conflicting conclusions may be encountered when a study examines the validity of a proposed tool in a healthcare setting or outcome different from those the tool was primarily developed for.

*Integration and Socio-Technical Context*

As is the case with other healthcare interventions, examining the post-implementation impact of predictive tools is challenging, since it is confounded by co-occurrent socio-technical factors [173-175]. This is complicated further by the fact that predictive tools are often integrated into electronic health record systems, since this facilitates their use, and are influenced by their usability [41, 176]. The usability, therefore, is an essential and major contributing factor in the wide acceptance and successful implementation of predictive tools and other CDS systems [41, 177]. It is clearly essential to involve user clinicians in the design and usability evaluations of predictive tools before their implementation. This should eliminate their concerns that integrating predictive tools into their workflow would increase their workload, consultation times, or decrease their efficiency and productivity [178].

Likewise, well designed post-implementation impact evaluation studies are required in order to explore the influence of organisational factors and local differences on the success or failure of predictive tools [179, 180]. Data availability, IT systems capabilities, and other human knowledge and organisational regulatory factors are crucial for the adoption, acceptance, and successful implementation of predictive tools. These factors and differences need to be included in the tools assessments, as they are important when making decisions about selecting predictive tools, in order to estimate the feasibility and resources needed to implement the tools. We have to acknowledge that it is not possible to include such wide range of variables in deciding or presenting the grades assigned by the framework to the predictive tools, which remain simply at a high-level. However, all the necessary information, technical specifications, and requirements of the tools, as reported in the published evidence, should be fully accessible to the users, through the framework detailed reports on the predictive tools. Users can compare such information, of one or more tools, to what they have at their healthcare settings, then make selection and implementation decisions.

*Local Data*

There is a challenging trade-off between the generalisability and the customisation of a given predictive tool. Some algorithms are developed using local data. For example, Bouvy's prognostic model, for mortality risk in patients with heart failure, uses life quality and expectancy scores from the Netherlands [181]. Similarly, Fine's prediction rule identifies low-risk patients with community-acquired pneumonia based on national rates of acquired infections in the United States [182]. This necessitates adjustment of the algorithm to the local context, therefore producing a new version of the tool, which requires re-evaluation.

*Other Considerations*

GRASP evaluates predictive tools based on the existing published evidence. Therefore, it is subject to publication bias, since statistically positive results are more likely to be published than negative or null results [183, 184]. The usability and potential effect of predictive tools are less studied and hence the published evidence needed to support level B of the grading system is often lacking. We have nevertheless chosen to keep this element in GRASP since it is an essential part of the safety evaluation of any healthcare technology. It also allows for early redesign and better workflow integration, which leads to higher utilisation rates [112, 155, 185]. By keeping it, we hope to encourage tool developers and evaluators to increase their execution and reporting of these type of studies.

The grade assigned to a tool provides relevant evidence-based information to guide choices on predictive tools for clinical decision support, but it is not prescriptive. An A1 tool is not always better than an A2 tool. A user may prefer an A2 tool showing improved patient safety in two observational studies rather than an A1 tool showing reduced cost in one experimental study. The grade is a code (not an ordinal quantity) that provides information on three relevant dimensions: phase of evaluation, level of evidence, and direction of evidence as reported in the literature.

*Study Limitations and Future Work*

We applied the GRASP framework to only five predictive tools and consulted a small number of healthcare experts for their feedback. This could have limited the conclusions about the framework's coverage and/or validity. Although GRASP framework is not a predictive tool, it could be thought of as a technology of Grade C3, since it has only been internally validated after development. However, conducting a large-scale validation study of the framework, extending the application of the framework to a larger number of predictive tools, and studying its effect on end-users' decisions is out of the scope of this study. To validate, test and improve the GRASP framework, the authors are currently working on three more studies. The first study should validate the design and content of the framework, through seeking the feedback of a wider international group of healthcare experts, who have published work on developing, implementing or evaluating predictive tools. This study should help to update the criteria used, by the framework, to grade predictive tools and improve the details provided, by the framework, to the end users. The second study should validate and evaluate the impact of using the framework on improving the decisions made by clinicians, regarding evaluating and choosing predictive tools. The experiment should compare the performance and outcomes of clinicians' decisions with and without using the framework. Finally, the third study aims to apply the framework to a larger consistent group of predictive tools, used for the same clinical task. This study should show how the framework provides clinicians with an evidence-based method to compare, evaluate and select predictive tools, through reporting and grading tools based on the critical appraisal of published evidence.

Implementing the GRASP framework as an online platform will enable clinicians and clinical practice guideline developers to access detailed information, reported evidence and grades of predictive tools. However, keeping such grading system up-to-date is a challenging task, since it requires updating tools' assessments and grades when new evidence becomes available. Therefore, a sustainable GRASP system will require employing automated or semi-automated methods for searching and processing new information. Finally, we recommend that GRASP framework be applied to predictive tools by working groups of professional organisations, in order to provide consistent results and increase reliability and credibility for end users. These professional organisations should also be responsible for making their associates aware of the availability of such evidence-based information on predictive tools, in a similar way of announcing and disseminating clinical practice guidelines.

**Abbreviations**

AUC: Area Under the Curve. CDS: Clinical Decision Support. CHALICE: Children's Head Injury Algorithm for the Prediction of Important Clinical Events. CHARMS: Critical Appraisal and Data Extraction for Systematic Reviews of Prediction Modelling Studies. CT: Computed Tomography. CTPA: Computed Tomography Pulmonary Angiography. ED: Emergency Department. E.g.: For example, GRADE: The Grading of Recommendations Assessment, Development and Evaluation. GRASP: Grading and Assessment of Predictive tools. ICU: Intensive Care Unit. I.e.: In other words or more precisely. IT: Information Technology. LACE: Length of Stay, Admission Acuity, Comorbidity and Emergency Department Visits. MDCalc: Medical Calculators. MEWS: Modified Early Warning Score. NASSS: Nonadoption, Abandonment, and Challenges to the Scale-Up, Spread, and Sustainability. PECARN: Paediatric Emergency Care Applied Research Network. RCTs: Randomised Controlled Trials. ROC = Receiver Operating Characteristic Curve. TRIPOD: Transparent Reporting of a Multivariable Prediction Model for Individual Prognosis or Diagnosis. UK: United Kingdom. US: United States.

**Declarations**

*Acknowledgments*

We would like to thank Professor William Runciman, Professor of Patient Safety, Dr Megan Sands, Senior Staff Specialist in Palliative Care, Dr Thilo Schuler, Radiation Oncologist, Dr Louise Wiles, Patient Safety Research Project Manager, and research fellows Dr Virginia Mumford and Dr Liliana Laranjo, for their contribution in testing the GRASP framework and for their valuable feedback and suggestions.

*Funding*

This work was supported by the Commonwealth Government Funded Research Training Program, Australia.

*Availability of data and materials*

Data sharing is not applicable to this article as no datasets were generated or analysed during the current study.

*Authors' contributions*

MK contributed to the conception and detailed design of the framework. BG and FM supervised the study from the scientific perspective. BG was responsible for the overall supervision of the work done, while FM was responsible for providing advice on the enhacement of the methodology used. All the authors have been involved in drafting the manuscript and revising it. Finally, all the authors gave approval of the manuscript to be published and agreed to be accountable for all aspects of the work.

*Ethics approval and consent to participate*

No ethics approval was required for any element of this study.

*Consent to publication*

Not applicable.

*Competing interests*

The authors declare that they have no competing interests.

*Author details*

[1]Australian Institute of Health Innovation, Faculty of Medicine and Health Sciences, Macquarie University, 75 Talavera Road, North Ryde, Sydney, NSW 2113, Australia.

**Figures**

**The Appendix**

*Phases, Criteria, and Measures of Evaluation*

Table 3: Phases, Criteria, and Measures of Evaluating Predictive Tools

| Phase of Evaluation | Criteria of Evaluation | Definitions | Example Measures* |
|---|---|---|---|
| Before Implementation | Predictive Performance | The ability of the predictive tool to utilise clinical variables and quantify relevant patient characteristics to produce an outcome that can be used to supports diagnostic, prognostic or therapeutic decisions made by clinicians and other healthcare professionals [9, 10]. | Discrimination:<br>• Sensitivity<br>• Specificity<br>• AUC, ROC, and C-Statistic<br>• D-Statistic<br>• Log-Rank Test.<br><br>Calibration:<br>• Calibration Plots & Curves<br>• Hosmer-Lemeshow test<br>• The Brier score. |
| During Implementation | Usability | The degree to which the predictive tool can be used by the specified users to achieve specified and quantifiable objectives in a specified context of use [106, 107]. | • Effectiveness of task management (accuracy and completeness).<br>• Efficiency of utilising resources.<br>• Users' satisfaction, comfort with, and positive attitudes towards, the use of the tools.<br>• Learnability<br>• Memorability<br>• Freedom of Errors. |
| | Potential Effect | The expected, estimated or calculated impact of using the tool on different healthcare aspects, processes or outcomes, assuming the tool has been successfully implemented and used in the clinical practice, as designed by its developers [40, 99]. | • Clinical Effectiveness (Clinical Patient Outcomes).<br>• Patient Safety (Complications, Side Effects, or Medical Errors).<br>• Healthcare Efficiency (Utilisation of Resources, Such as Time and Money). |
| After Implementation | Post-Implementation Impact | The achieved change or influence of a predictive tool on different healthcare aspects, processes or outcomes, after the tool has been successfully implemented and used in the clinical practice, as designed by its developers [2, 41]. | • Clinical Effectiveness (Clinical Patient Outcomes).<br>• Patient Safety (Complications, Side Effects, or Medical Errors).<br>• Healthcare Efficiency (Utilisation of Resources, Such as Time and Money). |

\* These measures of evaluation are examples, the list is not meant to be exhaustive; literature on predictive tools may evaluate them along other measures.

*Evaluating Evidence Direction*

Table 4: Evaluating Evidence Direction Based on the Conclusions of Studies

| Conclusions of Studies | | | Overall Direction of Evidence |
|---|---|---|---|
| **Positive *** | **Equivocal ** ** | **Negative *** ** | |
| ✓ | | | Positive |
| ✓ | | ✓ | Mixed |
| ✓ | ✓ | ✓ | |
| ✓ | ✓ | | |
| | | ✓ | Negative |
| | ✓ | ✓ | |
| | ✓ | | |
| * Positive Conclusion | • The tool shows positive valid predictive performance, usability, potential effect, or post-implementation impact, which are desirable and/or superior to other methods/tools, if the study includes a comparison. | | |
| ** Equivocal Conclusion | • The tool shows positive valid predictive performance or usability, which are acceptable, but not superior to other methods/tools, if the study includes a comparison.<br>• The tool does not show positive potential effect or post-implementation impact. These are inferior to other methods/tools, if the study includes a comparison. | | |
| *** Negative Conclusion | • The tool shows that predictive performance or usability is poor, not acceptable, or inferior to other methods/tools, if the study includes a comparison.<br>• The tool shows negative potential effect or post-implementation impact (leads to deterioration instead of improvement), whether in comparison or not. | | |

*GRASP Detailed Reports on Predictive Tools*

Table 5: LACE Index for Readmission – Grade C1

| Name | LACE Index for Readmission |
|---|---|
| Authors/Year | Dr. Carl van Walraven, Canada, 2010 |
| Intended use | Predicts 30 days readmission or death risk of medical and surgical inpatients after discharge |
| Intended user | Used by nurses at patient discharge |
| Category | Prognostic |
| Clinical area | All medical/surgical areas |
| Target Population | Hospitalised patients |
| Target Outcome | 30 days readmission or death |
| Action | Inform the clinical team about patients at high risk for readmission |
| Input source | Objective data (Data is available in the EHR – electronic health record, or manually obtained from the patient medical record). |
| Input type | Administrative data: Length of stay (days), Admission acuity (yes/no), Comorbidity (Charlson Index), Number of ED visits within 6 months. |
| Local context | Input does not depend on local context of data |
| Methodology | Multivariable logistic regression analysis |
| **Endorsement** | Recommended by:<br>• Texas Healthcare Association, USA.<br>• American Heart Association, USA.<br>• Michigan Care Management Resource Center, USA |
| **Automation Flag** | Manual |
| **Tool Citations** | 455 — Reported in 7 studies |

| Phase of Evaluation | Level of Evidence | Grade | Evaluation Studies |
|---|---|---|---|
| **Phase C: Before implementation Does the tool work? Is it possible?** | Internal validation | C3 | Developed and tested for internal validity:<br>• van Walraven et al, 2010 [123] |
| | External validation | C2 | Tested for externally validity:<br>• Gruneir et al, 2011 [128] |
| | External validation multiple times | C1 | Tested for external validity again:<br>• Au et al, 2012 [127]<br>Negative conclusion validation/performance studies:<br>• Cotter et al, 2012 [129]<br>• Wang et al, 2014 [130]<br>• Low et al, 2015 [131]<br>• Yu et al, 2015 [132] |
| **Phase B: During implementation: Is the tool practicable?** | Potential effect | B2 | Not reported |
| | Usability | B1 | Not reported |
| **Phase A: After implementation: Is the tool desirable?** | Evaluation of Post-Implementation Impact on Clinical Effectiveness, Patient Safety or Healthcare Efficiency | A3 | No subjective studies reported |
| | | A2 | No observational studies reported |
| | | A1 | No experimental studies reported |
| **Final Grade** | Grade C1 | | A1 · A2 · A3 · B1 · B2 · ◐ · C2 · ● |
| **Direction of Evidence** | ● Positive Evidence | | ◐ Mixed Evidence Supporting Positive Conclusion |
| | ○ Negative Evidence | | ◑ Mixed Evidence Supporting Negative Conclusion |
| **Justification** | LACE Index is a prognostic tool designed to predict 30 days readmission or death after discharge from hospital. It uses multivariable logistic regression analysis of administrative data: length of stay, admission acuity, comorbidity (Charlson Comorbidity Index) and emergency department (ED) visits in the last six months, to produce a risk score [123]. The tool has been tested for external validity twice: using a sample of 26,045 patients from six hospitals in Toronto and a sample of 59,652 patients from all hospitals in Alberta. The LACE Index showed external validity and superior predictive performance to previous tools endorsed by the Centres for Medicare and Medicaid Services [127, 128]. Two studies examined LACE Index predictive performance on small sub-population samples: 507 geriatric patients in the UK, and 253 congestive heart failure patients in the USA, and found that the index performed poorly [129, 130]. Two more studies reported that LACE Index works well but their own developed tools performed better [131, 132]. Using the mixed evidence protocol, the mixed evidence supports external validity, since the two negative conclusion studies have been conducted on very small samples of patients and a different subpopulation than the one LACE was developed for. There was no published evidence on the usability, potential effect or post-implementation impact of LACE Index. Accordingly, LACE Index has been assigned Grade C1. |

Table 6: Centor Score for Streptococcal Pharyngitis – Grade B1

| | |
|---|---|
| **Name** | Centor Score for Streptococcal Pharyngitis |
| **Authors/Year** | Dr. Robert M. Centor, USA, 1981. Modified later by Dr. Warren McIsaac, Canada, 1998. |
| **Intended use** | Estimate the probability that pharyngitis is streptococcal in adult patients presenting to the emergency department with sore throat |
| **Intended user** | Used by physicians at ED as part of the clinical examination |
| **Category** | Diagnostic |
| **Clinical area** | Infectious diseases |

| Target Population | Patients visiting the emergency department |
|---|---|
| Target Outcome | Streptococcal pharyngitis |
| Action | Consider rapid strep testing and/or culture |
| Input source | Objective data (clinical examination) + Subjective data (symptoms described by patient) |
| Input type | Clinical data: Age (3-14, 15-44 & >45 years), Exudate or swelling on tonsils (yes/no), Tender/swollen anterior cervical lymph nodes (yes/no), Temp >38°C (100.4°F) (yes/no), Cough (present/absent). Data is obtained from the patient. |
| Local context | Input does not depend on local context of data |
| Methodology | Rule-based algorithm |
| Endorsement | Recommended by:<br>• Department of Health, New South Wales, Australia<br>• American Academy of Family Physicians, United States<br>• The National Institute for Health and Care Excellence, United Kingdom |
| Automation Flag | Manual |
| Tool Citations | 715 — Reported in 15 studies |

| Phase of Evaluation | Level of Evidence | Grade | Evaluation Studies |
|---|---|---|---|
| **Phase C:** Before implementation Does the tool work? Is it possible? | Internal validation | C3 | Developed and tested for internal validity:<br>• Centor et al, 1981 [124] |
| | External validation | C2 | Tested for external validity:<br>• Wigton, Connor & Centor, 1986 [140] |
| | External validation multiple times | C1 | Tested for external validity multiple times:<br>• Poses et al, 1986 [139]<br>• Meland, Digranes & Skjærven, 1993 [138]<br>• Ebell et al, 2000 [135]<br>• McIsaac et al, 2004 [137]<br>• Aalbers et al, 2011 [133]<br>• Fine, Nizet & Mandl, 2012 [136]<br>• Alper et al, 2013 [134] |
| **Phase B:** During implementation: Is the tool practicable? | Potential effect | B2 | Not reported |
| | Usability | B1 | Reported usability testing is positive:<br>• Feldstein et al, 2017 [141] |
| **Phase A:** After implementation: Is the tool desirable? | Evaluation of Post-Implementation Impact on Clinical Effectiveness, Patient Safety or Healthcare Efficiency | A3 | No subjective studies reported |
| | | A2 | No observational studies reported |
| | | A1 | One RCT show positive post-implementation impact of Centor score on reducing unnecessary antibiotics prescribing:<br>• McIsaac et al, 1998 [142]<br>One observational study + 3 RCTs show negative conclusions (No impact of Centor score on antibiotics prescribing):<br>• McIsaac et al, 1998 [144]<br>• Poses, Cebul & Wigton, 1995 [145]<br>• Worrall et al, 2007 [146]<br>• Little et al, 2014 [143] |
| **Final Grade** | Grade B1 |  A2 A3  B2  C2  | |
| **Direction of Evidence** |  Positive Evidence |  Mixed Evidence Supporting Positive Conclusion | |
| |  Negative Evidence |  Mixed Evidence Supporting Negative Conclusion | |
| **Justification** | Centor score is a diagnostic tool that uses a rule-based algorithm on clinical data to estimate the probability that pharyngitis is streptococcal in adults who present to ED complaining of sore throat [124]. The score has been tested for external validity multiple times and all studies reported positive conclusions [133-140]. This qualifies Centor score for Grade C1. One study conducted a multicentre cluster RCT usability testing of the integration of Centor score into electronic health records. The study used "Think Aloud" testing with ten primary care providers, post interaction surveys in addition to screen captures and audio recordings to evaluate usability. Within the same study, another "Near Live" testing, with eight primary care providers, was conducted. Conclusions reported positive usability of the tool and positive feedback of users on the easiness of use and usefulness [141]. This qualifies Centor score for Grade B1. Evidence of the post-implementation impact of Centor score post-implementation is mixed. One RCT conducted in Canada reported a clinically important 22% reduction in overall antibiotic prescribing [142]. Four other studies, three of which were RCTs, reported that implementing Centor score did not reduce antibiotic prescribing in clinical practice [143-146]. Using | | | |

| | the mixed evidence protocol, the mixed evidence does not support positive post-implementation impact of Centor score. Therefore, Centor score has been assigned Grade B1. |

Table 7: Wells' Criteria for Pulmonary Embolism – Grade A2

| Name | Wells' Criteria for Pulmonary Embolism | | |
|---|---|---|---|
| Authors/Year | Dr. Phil Wells, Canada, 1998. | | |
| Intended use | Calculates the pre-test probability (risk)v of pulmonary embolism at the bedside without imaging | | |
| Intended user | Used by physicians at ED as part of the clinical examination | | |
| Category | Diagnostic | | |
| Clinical area | Cardiovascular diseases | | |
| Target Population | Patients visiting the emergency department | | |
| Target Outcome | Pulmonary embolism | | |
| Action | Rule out high risk patients with computed tomography angiography | | |
| Input source | Objective data (clinical examination) + Subjective data (symptoms described by patient). | | |
| Input type | Clinical data: Clinical signs and symptoms of DVT (yes/no), PE is #1 diagnosis OR equally likely (yes/no), Heart rate > 100 (yes/no), Immobilization at least 3 days OR surgery in the previous 4 weeks (yes/no), Previous, objectively diagnosed PE or DVT (yes/no), Haemoptysis (yes/no), Malignancy w/ treatment within 6 months or palliative (yes/no). | | |
| Local context | Input does not depend on local context of data | | |
| Methodology | Rule-based algorithm | | |
| Endorsement | Recommended by:<br>• New South Wales Agency for Clinical Innovation, Australia<br>• The Royal Australian College of General Practitioners, Australia | | |
| Automation Flag | Manual | | |
| Tool Citations | 1,260 | Reported in 13 studies | |
| Phase of Evaluation | Level of Evidence | Grade | Evaluation Studies |
| Phase C: Before implementation Does the tool work? Is it possible? | Internal validation | C3 | Developed and tested for internal validity:<br>• Wells et al, 1998 [122]<br>• Wells et al, 2000 [121]<br>• Wells et al, 2001 [125] |
| | External validation | C2 | Tested for external validity:<br>• Page, 2006 [149] |
| | External validation multiple times | C1 | Tested for external validity multiple times:<br>• Gibson et al, 2008 [148]<br>• Klok et al, 2008 [153]<br>• Söderberg et al, 2009 [151]<br>• Geersing et al, 2012 [147]<br>• Arslan et al, 2013 [152]<br>• Posadas-Martínez et al, 2014 [150]<br>• Turan et al, 2017 [154] |
| Phase B: During implementation: Is the tool practicable? | Potential effect | B2 | Not reported |
| | Usability | B1 | Reported usability testing is positive:<br>• Press et al, 2015 [155] |
| Phase A: After implementation: Is the tool desirable? | Evaluation of Post-Implementation Impact on Clinical Effectiveness, Patient Safety or Healthcare Efficiency | A3 | No subjective studies reported |
| | | A2 | Observational before-and-after intervention study showing positive post-implementation impact of Wells' Criteria on healthcare efficiency:<br>• Murthy et al, 2016 [156] |
| | | A1 | No experimental studies reported |
| Final Grade | Grade A2 | | A1 ● A3 ● B2 ● C2 ● |
| | ● Positive Evidence | | ◐ Mixed Evidence Supporting Positive Conclusion |

| Direction of Evidence | ◯ Negative Evidence | ◐ Mixed Evidence Supporting Negative Conclusion |
|---|---|---|
| Justification | Wells' criteria is a diagnostic tool used in ED to estimate pre-test probability of pulmonary embolism [121, 122]. Using a rule-based algorithm on clinical data, the tool calculates a score that excludes pulmonary embolism without diagnostic imaging [125]. The tool was tested for external validity multiple times [147-151] and its predictive performance has been also compared to other predictive tools [152-154]. In all studies, Wells' criteria was reported valid, which qualifies it for Grade C1. One study conducted usability testing for the integration of the tool into the electronic health record system of a tertiary care centre's ED. The study identified a strong desire for the tool and received positive feedback on the usefulness of the tool itself. Subjects responded that they felt the tool was helpful, organized, and did not compromise clinical judgment [155]. This qualifies Wells' criteria for Grade B1. The post-implementation impact of Well's Criteria on efficiency of computed tomography pulmonary angiography (CTPA) utilisation has been evaluated through an observational before-and-after intervention study. It was found that the Well's Criteria significantly increased the efficiency of CTPA utilisation and decreased the proportion of inappropriate scans [156]. Therefore, Well's Criteria has been assigned Grade A2. | |

Table 8: Modified Early Warning Score (MEWS) – Grade A2

| Name | Modified Early Warning Score (MEWS) for Clinical Deterioration | | |
|---|---|---|---|
| Authors/Year | Dr. Christian Peter Subbe, UK, 2001 | | |
| Intended use | Early detection of inpatients' clinical deterioration, calculate chance of ICU admission or death within 60 days and potential need for higher levels of care. | | |
| Intended user | Used by nurses at bedside | | |
| Category | Prognostic | | |
| Clinical area | General Medicine | | |
| Target Population | Hospitalised patients | | |
| Target Outcome | Clinical deterioration/death | | |
| Action | Consider higher level of care for patient (e.g. transfer to ICU) | | |
| Input source | Objective (Data from EHR – electronic health record) | | |
| Input type | Clinical data: Systolic BP, Heart rate, Respiratory rate, Temperature, AVPU Score. | | |
| Local context | Input does not depend on local context of data | | |
| Methodology | Rule-based algorithm | | |
| Endorsement | Recommended by:<br>• Australian Commission on Safety and Quality in Health Care, Australia<br>• National Health Services, United Kingdom | | |
| Automation Flag | Automated (However, in some hospitals a manual version is still used by nurses) | | |
| Tool Citations | 1,176 | Reported in 13 studies | |
| Phase of Evaluation | Level of Evidence | Grade | Evaluation Studies |
| Phase C: Before implementation Does the tool work? Is it possible? | Internal validation | C3 | Developed and tested for internal validity:<br>• Subbe et al, 2001 [126] |
| | External validation | C2 | Tested for External validity:<br>• Armagan et al, 2008 [157] |
| | External validation multiple times | C1 | Tested for external validity multiple times:<br>• Burch, Tarr & Morroni, 2008 [158]<br>• Dundar et al, 2016 [159]<br>• Gardner-Thorpe et al, 2006 [160]<br>• TANRIÖVER et al, 2016 [162]<br>• Wang et al, 2016 [163]<br>• Salottolo et al, 2017 [161]<br>One negative conclusion validation/performance study:<br>• Tirotta et al, 2017 [164] |
| Phase B: During implementation: Is the tool practicable? | Potential effect | B2 | Not reported |
| | Usability | B1 | Not reported |

| Phase A: After implementation: Is the tool desirable? | Evaluation of Post-Implementation Impact on Clinical Effectiveness, Patient Safety or Healthcare Efficiency | A3 | No subjective studies reported |
|---|---|---|---|
| | | A2 | One observational before-and-after intervention study failed to prove positive post-implementation impact of the MEWS on patient safety:<br>• Subbe et al, 2003 [165]<br>Three observational before-and-after intervention studies showed positive post-implementation impact of the MEWS on patient safety:<br>• Moon et al, 2011 [168]<br>• De Meester et al, 2013 [166]<br>• Hammond et al, 2013 [167] |
| | | A1 | No experimental studies reported |
| **Tool Grade** | Grade A2 | A1 ◐  A3   B1   B2 ◐  C2 ● | |
| **Direction of Evidence** | ● Positive Evidence | ◐ Mixed Evidence Supporting Positive Conclusion | |
| | ○ Negative Evidence | ◔ Mixed Evidence Supporting Negative Conclusion | |
| **Justification** | The MEWS is a prognostic tool for early detection of inpatients' clinical deterioration and potential need for higher levels of care. The tool uses a rule-based algorithm on clinical data to calculate a risk score [126]. The MEWS has been tested for external validity multiple times in different clinical areas, settings and populations [157-163]. All studies reported the tool is externally valid. However, one study reported MEWS poorly predicted the in-hospital mortality risk of patients with sepsis [164]. Using the mixed evidence protocol, the mixed evidence supports external validity, qualifying MEWS for Grade C1. No literature has been found regarding its usability or potential effect. The MEWS has been implemented in different healthcare settings. One observational before-and-after intervention study failed to prove positive post-implementation impact of the MEWS on patient safety in acute medical admissions [165]. However, three more recent observational before-and-after intervention studies reported positive post-implementation impact of the MEWS on patient safety. One study reported significant increase in frequency of patient observation and decrease in serious adverse events after intensive care unit (ICU) discharge [166]. The second reported significant increase in frequency of vital signs recording, 24h post-ICU discharge and 24h preceding unplanned ICU admission [167]. The third, an eight years study, reported that the post-implementation four years showed significant reductions in the incidence of cardiac arrests, the proportion of patients admitted to ICU and their in-hospital mortality [168]. Using the mixed evidence protocol, the mixed evidence supports positive post-implementation impact. The MEWS has been assigned Grade A2. | | |

Table 9: Ottawa Knee Rule – Grade A1

| Name | Ottawa Knee Rule |
|---|---|
| Authors/Year | Dr. Ian Stiell, Canada, 1995 |
| Intended use | Exclude the need for an X-ray for possible bone fracture in adult patients |
| Intended user | Used by emergency physicians as part of the clinical examination |
| Category | Diagnostic |
| Clinical area | Orthopaedics |
| Target Population | Patients visiting the emergency department |
| Target Outcome | Bone fracture |
| Action | Refer patient to knee imaging |
| Input source | Objective data (clinical examination) + Subjective data (symptoms described by patient) |
| Input type | Clinical data: Age ≥55 (yes/no), Isolated tenderness of the patella (no other bony tenderness) (yes/no), Tenderness at the fibular head (yes/no), Unable to flex knee to 90° (yes/no), Unable to bear weight both immediately and in ED (4 steps, limping is okay) (yes/no). Data is obtained from the patient. |
| Local context | Input does not depend on local context of data |
| Methodology | Set of rules |
| Endorsement | Recommended by:<br>• Department of Emergency Medicine, Faculty of medicine, Ottawa University, Canada<br>• The Royal College of Radiologists, United Kingdom<br>• The National Institute for Health and Care Excellence, United Kingdom |
| Automation Flag | Manual |

| Tool Citations | 227 | | Reported in 15 studies | |
|---|---|---|---|---|
| Phase of Evaluation | Level of Evidence | Grade | Evaluation Studies | |
| Phase C: Before implementation Does the tool work? Is it possible? | Internal validation | C3 | Developed and tested for internal validity:<br>• Stiell et al, 1995 [120] | |
| | External validation | C2 | Tested for externally validity | |
| | External validation multiple times | C1 | Externally tested for externally validity (One systematic review reported 11 validation studies):<br>• Bachmann et al, 2004 [169] | |
| Phase B: During implementation: Is the tool practicable? | Potential effect | B2 | Not reported | |
| | Usability | B1 | Not reported | |
| Phase A: After implementation: Is the tool desirable? | Evaluation of Post-Implementation Impact on Clinical Effectiveness, Patient Safety or Healthcare Efficiency | A3 | No subjective studies reported | |
| | | A2 | No observational studies reported | |
| | | A1 | Two nonrandomised controlled studies reported positive post-implementation impact of Ottawa knee rule on healthcare efficiency:<br>• Stiell et al, 1997 [170]<br>• Nichol et al, 1999 [171] | |
| Final Grade | Grade A1 | | ● A2 A3 B1 B2 ● C2 ● | |
| Direction of Evidence | ● Positive Evidence | | ◐ Mixed Evidence Supporting Positive Conclusion | |
| | ○ Negative Evidence | | ◒ Mixed Evidence Supporting Negative Conclusion | |
| Justification | Ottawa knee rule is a diagnostic tool used to exclude the need for an X-ray for possible bone fracture in patients presenting to the ED, using a simple five items manual check list [120]. It is one of the oldest, most accepted and successfully used rules in CDS. The tool has been tested for external validity multiple times. One systematic review identified 11 studies, 6 of them involved 4,249 adult patients and were appropriate for pooled analysis, showing high sensitivity and specificity [169]. Furthermore, two studies discussed the impact of implementing Ottawa knee rule on healthcare efficiency. One nonrandomised controlled trial with before-after and concurrent controls included a total of 3,907 patients seen during two 12-month periods before and after the intervention. The study reported that the rule decreased the use of knee radiography without patient dissatisfaction or missed fractures and was associated with reduced waiting times and costs per patient [170]. Another nonrandomised controlled trial reported that the proportion of ED patients referred for knee radiography was reduced. The study also reported that the practice based on the rule was associated with significant cost savings [171]. The Ottawa knee rule has been assigned Grade A1. | | | |

*Predictive Performance, Usability and Post-implementation Impact Tables of Predictive Tools*

Table 10: Predictive Performance of the Five Tools – Before Implementation

| Tool | Discrimination | | Calibration |
|---|---|---|---|
| | AUC/C-Statistic | Sensitivity, Specificity, Cut-Off | Hosmer–Lemeshow goodness-of-fit |
| LACE Index | • 0.68 (95% CI, 0.68–0.69) [123]<br>• 0.68 [127]<br>• 0.56 (95% CI, 0.46–0.66) [186] | • 66.3%, 53.3%, 50% [131] | • 14.1 (P=0.59) [123] |
| Centor Score | • 0.78 [124]<br>• 0.72 [136]<br>• 0.84 [134] | • 90%, 92%, 50% [124]<br>• 49%, 82%, 50% [133]<br>• 92%, 73%, 50% [137]<br>• 92%, 63%, 50% [134] | • Not reported |
| Wells' Criteria | • 0.71 [151]<br>• 0.75 [152]<br>• 0.79 (95% CI, 0.75-0.82) [150]<br>• 0.79 (95% CI, 0.72–0.87) [153]<br>• 0.76 [154]<br>• 0.74 (95% CI,0.72-0.76) [148] | • 83%, 48%, 50% [151]<br>• 65%, 81%, 50% [150]<br>• 100%, 56%, 50% [154]<br>• 95%, 51%, 50% [147] | • Not reported |
| MEWS | • 0.73 (95% CI, 0.69–0.77) – Hospitalisation [159]<br>• 0.89 (95% CI 0.84–0.94) – In-hospital mortality [159]<br>• 0.79 (95% CI, 0.74-0.83) – Mortality [161]<br>• 0.56 (95% CI, 0.51 to 0.62) – ICU Admission [161]<br>• 0.85 (95% CI, 0.77–0.91) [162]<br>• 0.80 (95% CI, 0.72–0.88) [36]<br>• 0.76 – Mortality [187] | • 88%, 68%, 50% (MEWS≥3) [160]<br>• 53%, 91%, 50% (MEWS≥4) Mortality [161]<br>• 17%, 94%, 50% (MEWS≥4) ICU Admission [161]<br>• 86%, 94%, 50% (MEWS≥4) [162]<br>• 75%, 83%, 50% (MEWS≥4) [188]<br>• 57%, 86%, 50% (MEWS≥4) [187] | • P=0.06 [189] |
| Ottawa Knee Rule | • Not reported | • 98.5%, 48.6%, 50% [169]*<br>• 100% [170]<br>• 100%, 42.8%, 50% [190]<br>• 95%, 44%, 50% [191] | • Not reported |

* A systematic review study.

Table 11: Usability of Two Predictive Tools – During Implementation

| Tool | Study Type | Method | Outcomes |
|---|---|---|---|
| Centor Score | Usability testing [141] | Think Aloud + Near Live | • Positive usability & feedback of users<br>  o Easiness of use<br>  o Usefulness |
| Wells' Criteria | Usability testing [155] | Think Aloud + Near Live | • Positive usability & feedback of users<br>  o Tool is helpful<br>  o Organized<br>  o Did not compromise clinical judgment |

Table 12: Post-Implementation Impact of Three Predictive Tools

| Tool | Study Type | Study Settings | Outcome | Effect Size |
|---|---|---|---|---|
| Wells' Criteria | Prospective before-and-after intervention study [156] | Public-sector tertiary-level and referral teaching hospital in South Africa | Efficiency of CTPA utilisation | 17.4% vs 30.7% (p=0.036) |
| | | | Inappropriate CTPA scans | 82.6% vs 69.3% (p=0.015) |
| MEWS | Prospective before-and-after intervention study [166] | A University Hospital, in Belgium | Frequency of patient observation | 0.99 vs 1.07 (p=0.005) |
| | | | Serious adverse events after ICU discharge | 5.7% vs 3.5% |
| | Prospective before-and-after intervention study [167] | The department of intensive care medicine, at a tertiary referral hospital in Brisbane, Australia | Vital signs documentation after ICU discharge | 210% (95% CI 148, 288%, p <0.001). |
| | | | Vital signs documentation before unplanned ICU admissions | 44% (95% CI, 3, 102%, p = 0.035). |
| | Retrospective analysis of prospectively collected data before-and-after intervention study [168] | The department of perioperative and critical care at a university teaching hospital in the United Kingdom | Cardiac arrest calls | 0.2% vs 0.4% (p<0.0001) |
| | | | Patients admitted to ICU | 2% vs 3% (p=0.004) |
| | | | In-hospital mortality of cardiac arrest patients | 42% vs 52% (p=0.05) |
| Ottawa Knee Rule | Nonrandomised controlled trial with before-after & concurrent controls [170] | The Emergency departments of two teaching and two community hospitals in Canada | Reduced time spent by patient | 85.7 minutes vs 118.8 minutes |
| | | | Cost savings per patient | US $80 vs US $183 |
| | Nonrandomised controlled trial with before-after & concurrent controls [171] | The Emergency departments of an academic and a community hospital in Canada. | Reduced proportion of knee injury patients referred to radiology | 77.6% vs 57.1% |
| | | | Cost savings per patient | $31 (95% CI, 22 to 44) to $34 (95% CI, 24 to 47). |

*Study Selection Process*

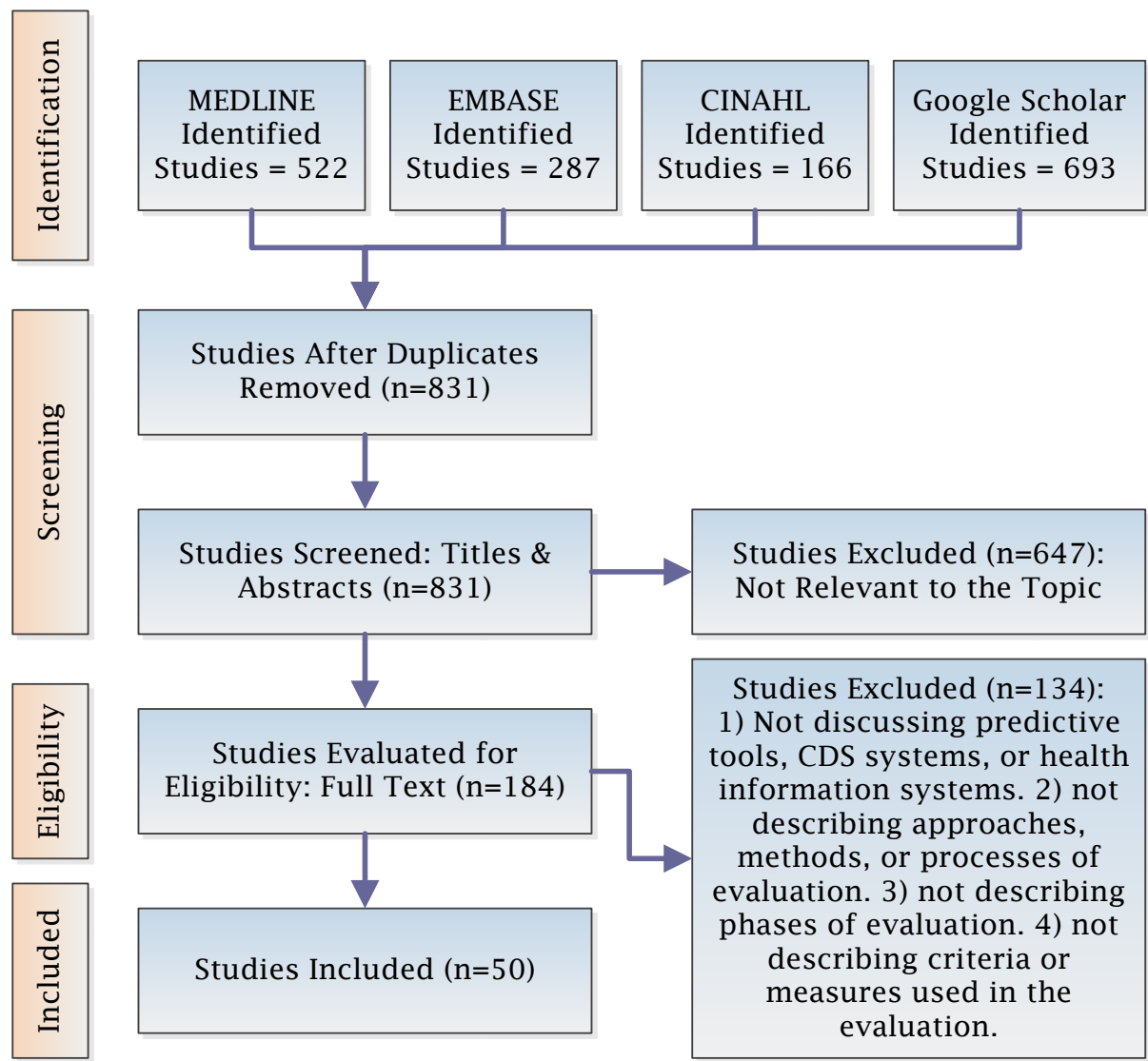

Figure 3: Study Selection for the Focused Review of Literature

*Searching the Literature for Published Evidence on Predictive Tools*

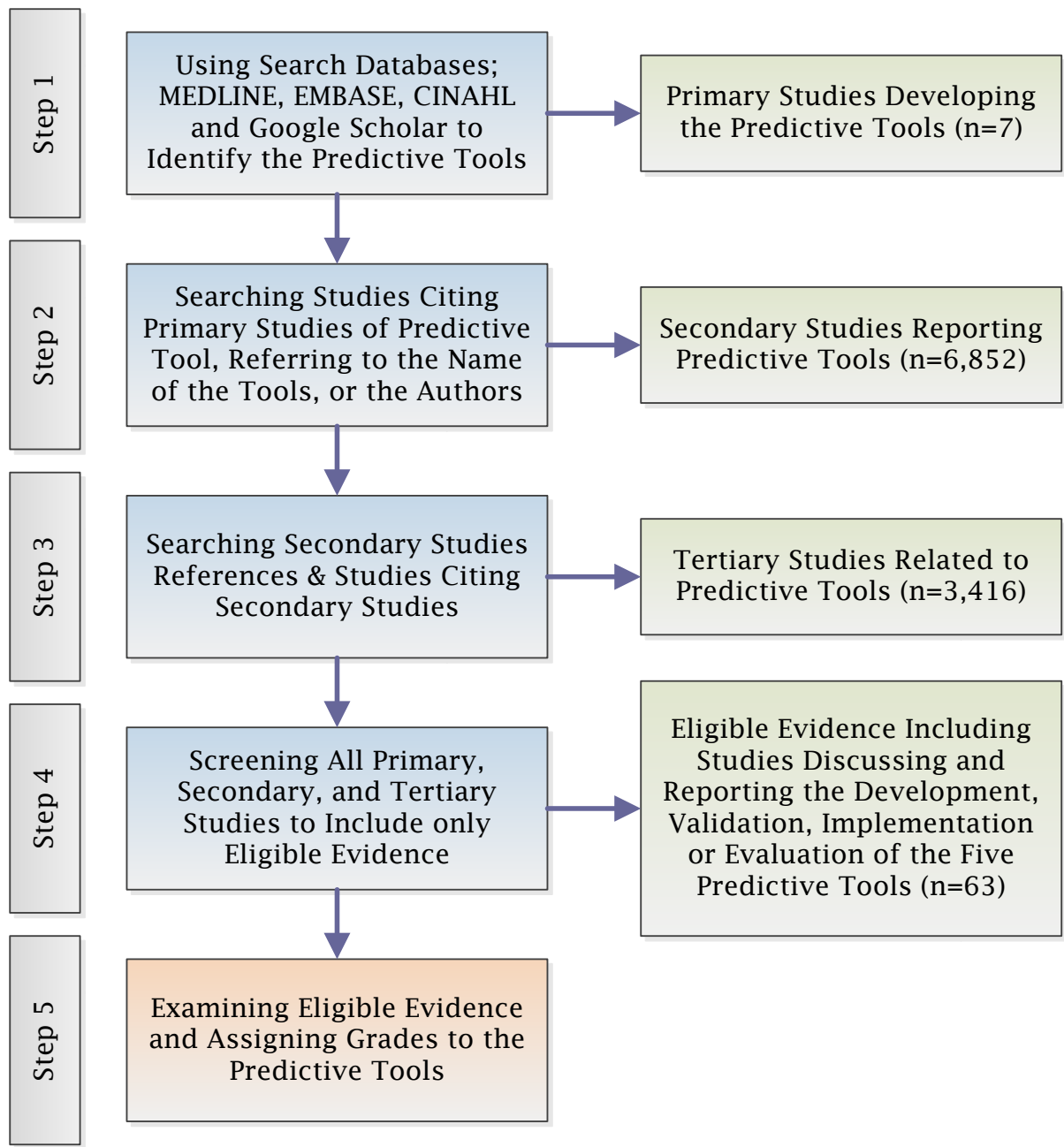

Figure 4: Searching the Literature for Published Evidence on Predictive Tools

*The Mixed Evidence Protocol*

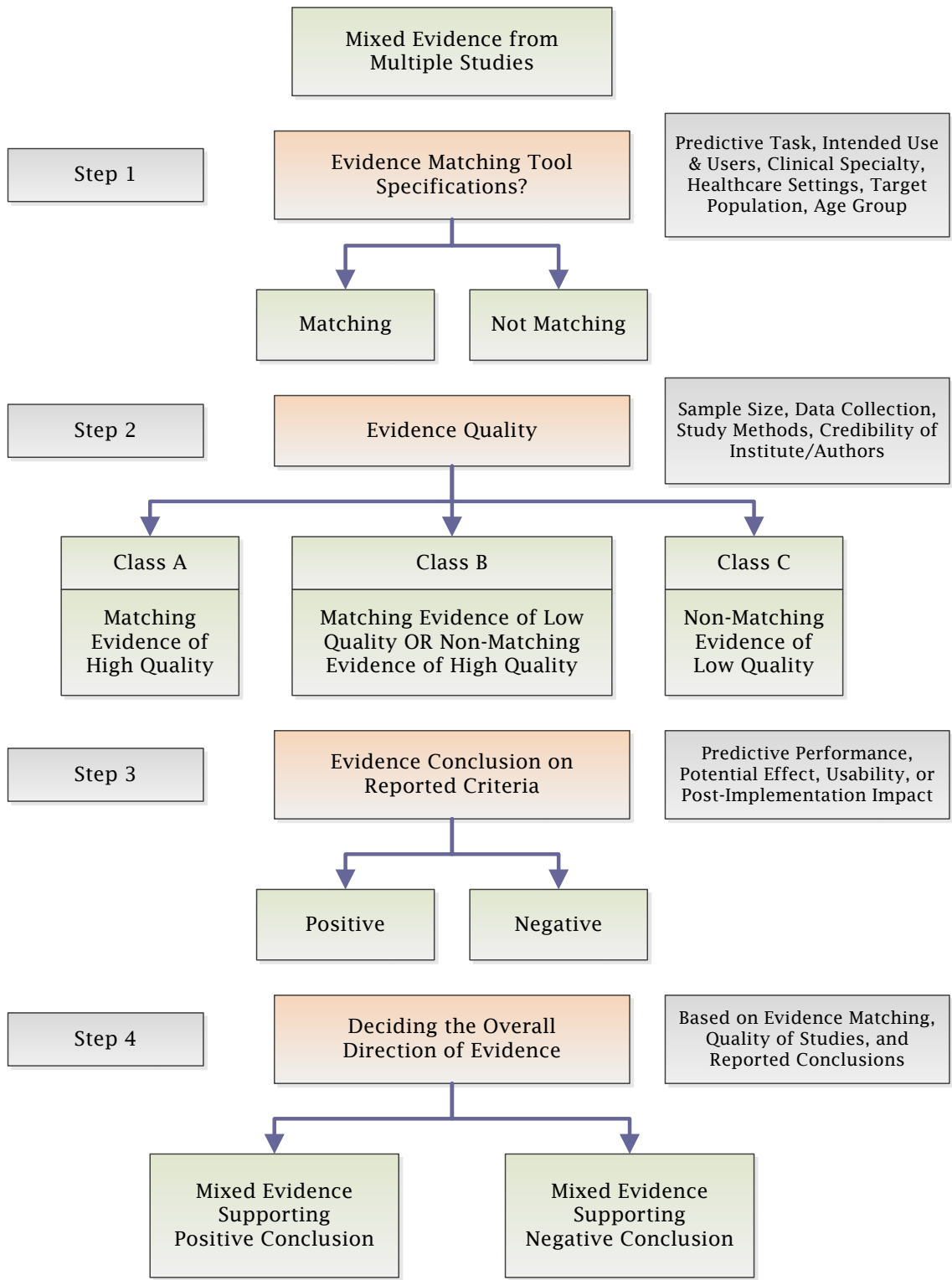

Figure 5: The Mixed Evidence Protocol

The mixed evidence protocol is based on four steps. Firstly, it considers the degree of matching between the evaluation study conditions and the original tool specifications, in terms of the predictive task, outcome, intended use and users, clinical specialty, healthcare settings, target population, and age group. Secondly, it considers the quality of the study, in terms of sample size, data collection, study methods, and credibility of institute or authors. Based on these two criteria, the studies in the mixed evidence on the tool are classified into 1) Class A: matching evidence of high quality, 2) Class B: matching evidence of low quality or non-matching evidence of high quality, and 3) Class C: non-matching evidence of low quality. Thirdly, it considers the evidence conclusion on the reported evaluation criteria; the predictive performance, potential effect, usability, and post-implementation impact. In the fourth step, studies evaluating predictive tools in closely matching conditions to the tool specifications and providing high quality evidence, Class A, are considered first; taking into account their conclusions on the evaluation criteria in deciding the overall direction of evidence. On the other hand, studies evaluating predictive tools in different conditions to the tool specifications and providing low quality evidence, Class C, are considered last. The conclusion of one study in Class A is considered a stronger evidence than the conflicting conclusions of any number of studies in Class B or C, and the overall direction of the evidence is decided towards the conclusion of the study of Class A. When multiple studies of the same class; for example Class A, report conflicting conclusions, then we compare the number of studies reporting positive conclusions to those reporting negative conclusions and the overall direction of the evidence is decided towards the conclusions of the larger group. If the two groups are of the same size, then we check if there are more studies in other classes, if not then we examine the reported evaluation criteria and their values in the two groups of studies.

*Performance Figures of Predictive Tools*

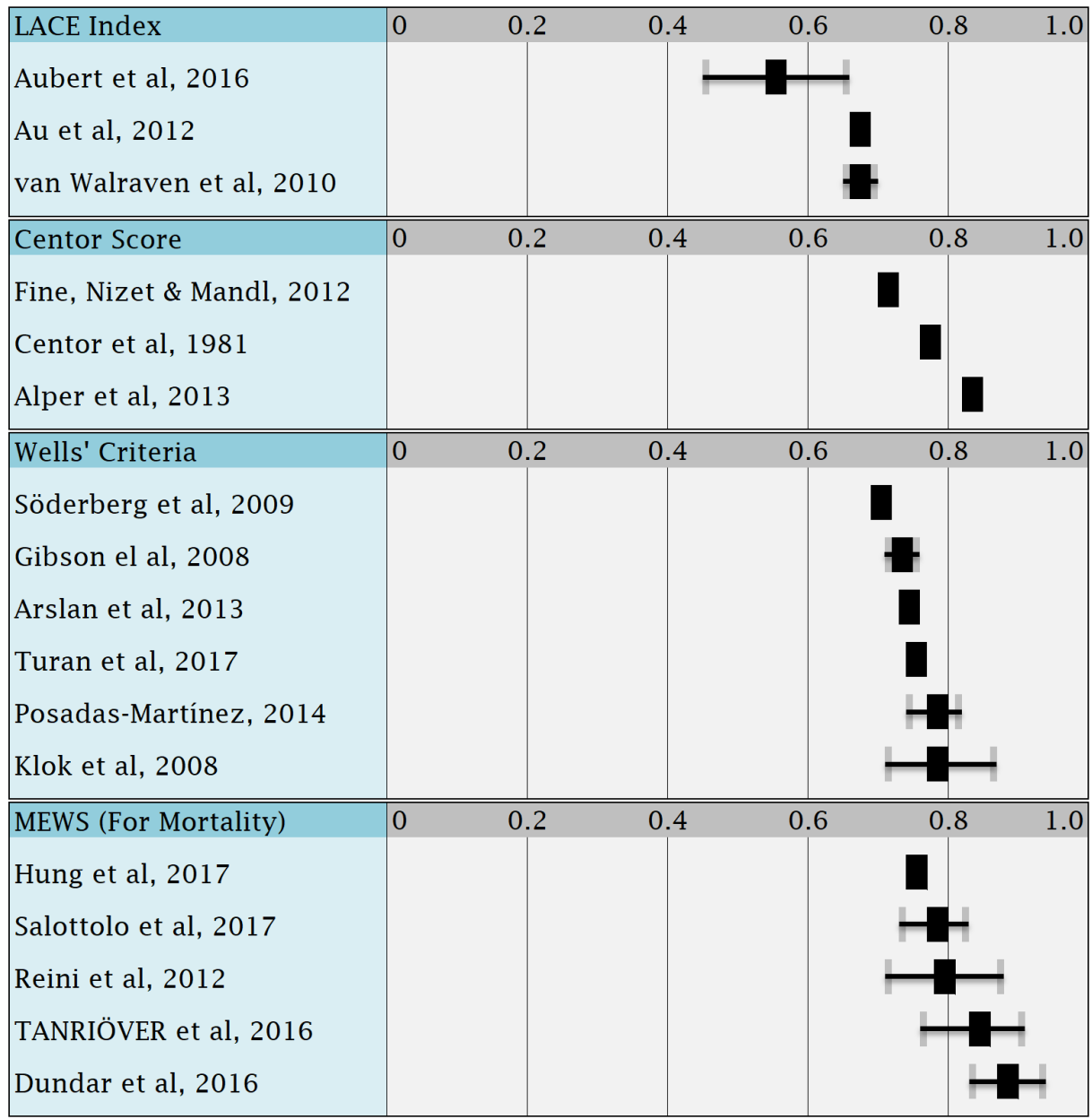

Figure 6: Reported C-Statistic of LACE Index, Centor Score, Wells Criteria and MEWS